\documentclass[conference]{IEEEtran}
\IEEEoverridecommandlockouts
\usepackage{cite}
\usepackage{multirow,tabulary}
\usepackage{algorithmic}
\usepackage{graphicx}
\usepackage{textcomp}
\usepackage{xcolor}
\usepackage{color}
\usepackage{subcaption}

\def\BibTeX{{\rm B\kern-.05em{\sc i\kern-.025em b}\kern-.08em
    T\kern-.1667em\lower.7ex\hbox{E}\kern-.125emX}}
\begin{document}

\title{Hierarchical Roofline Performance Analysis for Deep Learning Applications}
%Roofline Performance Analysis for DeepCAM - an MLPerf HPC Benchmark

\author{\IEEEauthorblockN{Charlene Yang, Yunsong Wang\\ Steven Farrell}
\IEEEauthorblockA{\textit{National Energy Research Scientific Computing Center} \\
\textit{Lawrence Berkeley National Laboratory}\\
Berkeley, CA USA\\
\{cjyang, yunsongwang, sfarrell\}@lbl.gov}
\and
\IEEEauthorblockN{Thorsten Kurth}
\IEEEauthorblockA{\textit{NVIDIA Corporation} \\
%\textit{2788 San Tomas Expressway}\\
Santa Clara, CA USA\\
tkurth@nvidia.com}
\and
\IEEEauthorblockN{Samuel Williams}
\IEEEauthorblockA{\textit{Computational Research Division} \\
\textit{Lawrence Berkeley National Laboratory}\\
Berkeley, CA USA\\
swwilliams@lbl.gov}
}

\maketitle

\begin{abstract}
This paper presents a practical methodology for collecting performance data necessary to conduct hierarchical Roofline analysis on NVIDIA GPUs.
It discusses the extension of the Empirical Roofline Toolkit for broader support of a range of data precisions and Tensor Core support and introduces a Nsight Compute based method to accurately collect application performance information. 
This methodology allows for automated machine characterization and application characterization for Roofline analysis across the entire memory hierarchy on NVIDIA GPUs, and it is validated by a complex deep learning application used for climate image segmentation.
We use two versions of the code, in TensorFlow and PyTorch respectively, to demonstrate the use and effectiveness of this methodology.
We highlight how the application utilizes the compute and memory capabilities on the GPU and how the implementation and performance differ in two deep learning frameworks.
\end{abstract}


%============================================================
\section{Introduction}

The Roofline model~\cite{williams2009roofline} is an intuitive performance model that can offer valuable insights into application performance, performance bottlenecks, and possible optimization opportunities.
Its capability to extract the key computational characteristics and abstract away the complexity of modern computer architectures has gained its popularity in recent years in both traditional high-performance computing (HPC) and machine learning.
Roofline is a throughput-oriented model centered around the interplay of computational capabilities, memory bandwidth, and data locality.  
Data locality is expressed as the arithmetic intensity (AI), the reuse of data once it is being loaded from memory, and it is commonly calculated as the ratio of the floating-point operations performed to the data movement, i.e. FLOPs per byte.
The sustained performance (GFLOP/s) is then bound by two terms:
\begin{equation}
\mbox{GFLOP/s} \leq \mbox{min} \left\{ 
  \begin{array}{@{}l@{}}
    \mbox{Peak GFLOP/s}							\\
    \mbox{Peak GB/s} \times \mbox{Arithmetic Intensity}	
  \end{array}
\right.
\label{eq:roofline_naive}
\end{equation}

The Roofline model conventionally only focuses on one level in the memory hierarchy, but this has been extended in recent years to the full memory system to help understand cache reuse and data locality and provide additional insights into code performance.
To facilitate the Roofline study, many tools and workflows have sprung to life,
for example, the Empirical Roofline Toolkit (ERT) developed at the Lawrence Berkeley National Laboratory, for more accurate machine characterization~\cite{ert2020,yang2018empirical}, and other tools, methodologies, and workflows for more streamlined application performance data collection in~\cite{nerscroofline,yang2018toward,madsen2020timemory,yang2019hierarchical}.
A range of studies have also been conducted on the application of Roofline in both traditional HPC~\cite{doerfler2016applying,koskela2018novel,yang2019hierarchical,del2020accelerating,gayatri2018case,arxiv-paper} and Machine Learning~\cite{yang2019hierarchical,javed2019performance,wang2020dlonsc}, and the extension and refinement of the model to other related topics such as instruction Roofline~\cite{ding2019instruction}, time-based Roofline~\cite{wang2020dlonsc}, Roofline scaling trajectories~\cite{ibrahim2019performance}, performance portability analysis based on Roofline~\cite{yang2018empirical}, and power and energy Roofline~\cite{powerroofline,alexpowerroofline}.

Deep learning has become one of the most dominant tools in areas such as pattern recognition, object detection, image segmentation, and language processing~\cite{lecun1995convolutional,goodfellow2014generative}, and its training or inference process usually takes a long time and requires significant computational resources.
To tackle this problem, many innovative methods have been proposed~\cite{kurth2018exascale,joubert2018attacking} to scale up such applications, and in this paper, we will focus on the Roofline-based performance modeling to analyze and examine how well various deep learning frameworks are utilizing the different aspects of the computer architecture, especially NVIDIA GPUs.

We will propose a practical methodology for collecting necessary performance data to conduct hierarchical Roofline analysis on NVIDIA GPUs.
There are two components to this methodology, machine characterization using the Empirical Roofline Toolkit (ERT)~\cite{ert2020} and application characterization using \texttt{Nsight Compute}~\cite{nsight2020}.
We will discuss the extension of ERT for support on multiple data precisions and Tensor Core operations, and the \texttt{Nsight Compute} metrics used to measure application performance such as the run time, sustained throughput, and data movement across the entire memory hierarchy.
This methodology then will be validated by a state-of-the-art deep learning application, DeepCAM~\cite{kurth2018exascale}, in climate image segmentation, to demonstrate its effectiveness in application analysis.
Two versions of the code will be examined, in TensorFlow and PyTorch respectively, and some insights will be highlighted on how deep learning applications, in general, utilize the compute/memory capabilities on NVIDIA GPUs and how the two deep learning frameworks, TensorFlow and PyTorch, can differ in implementation and performance. 

\section{Methodologies}

In this section, we will discuss the extension work done on the Empirical Roofline Toolkit (ERT) in order to support multiple data precisions (such as FP16) and Tensor Core operations on NVIDIA GPUs, and the set of metrics in \texttt{Nsight Compute} that can be used to measure application performance such as run time, sustained throughput and data movement at different levels of the memory hierarchy.
These two components together comprise the complete data collection methodology for machine and application characterization in a hierarchical Roofline analysis on NVIDIA GPUs.

%============================================================
\subsection{ERT Extensions for Machine Characterization}

The Empirical Roofline Toolkit (ERT)~\cite{ert2020} is developed and maintained by the Lawrence Berkeley National Laboratory. 
It consists of micro-kernels that are finely tuned to test the various aspects of computer architecture such as memory bandwidth and compute throughput. 
Compared to theoretical values or marketing numbers from vendors, this provides a more accurate understanding of the architecture's capability in real programming environments with real power, thermal constraints, and programming models.  

ERT is essentially a Python script that wraps around a range of micro-kernels written in C++ and parallelized with various programming models on different architectures. 
For example, OpenMP and MPI are used on Intel CPUs, CUDA is used on NVIDIA GPUs, and more micro-kernels are currently being added to support AMD architectures, IBM Power processors, and Intel GPUs. 
These micro-kernels are specifically tuned to test different aspects of the architecture and provide an upper bound for real-life applications on them, i.e. if such kernels can not reach certain performance, there is almost no hope for large complex applications in real life to achieve it.
\begin{figure}[!htp]
\centering
\includegraphics[width=\columnwidth]{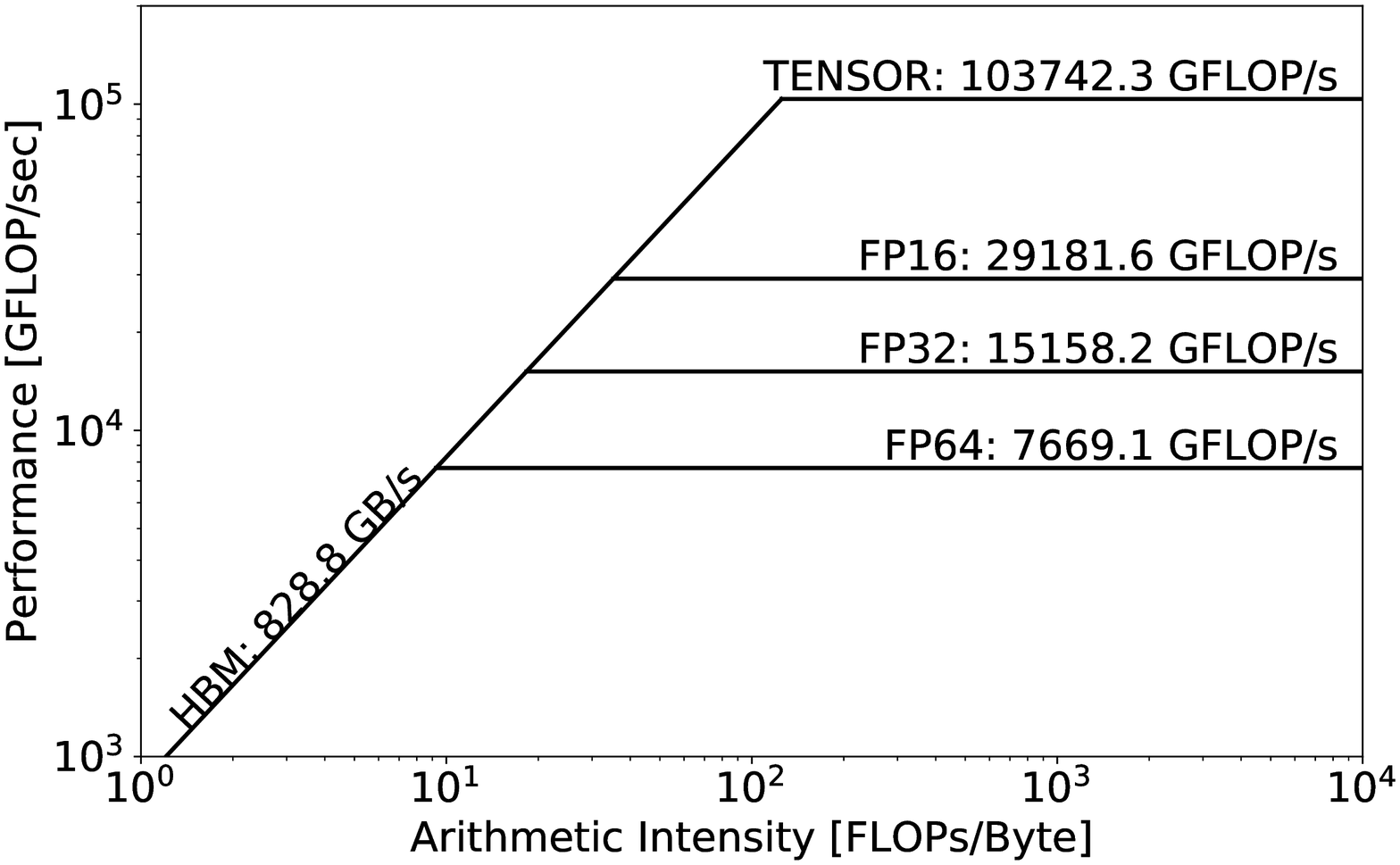}
\caption{Roofline graph generated by empirical results for V100 GPU including the new reduced-precision and tensor core ceilings.}
\label{fig:ert_roofline}
\end{figure}

The ERT prior to this paper only supports double precision (FP64) performance characterization and in this section, we will detail how we have extended it to support single-precision (FP32), and half-precision (FP16), as well as Tensor Core operations on NVIDIA GPUs. 
The resultant Roofline ceilings are shown in Fig.~\ref{fig:ert_roofline}, with 7.7~TFLOP/s for FP64, 15.2~TFLOP/s for FP32, 29.2~TFLOP/s for FP16 on the CUDA core, and 103.7~TFLOP/s on the Tensor Core, on V100 GPUs.

%\fix{do we have L2, L1 bandwidth measurement? is everything at 1312 MHz frequency?}

% The original ERT only supports double-precision performance measurement. In the rest of this section, we will introduce our latest work on implementing the measurement of single-precision (FP32), half-precision (FP16), and Tensor Core peak in ERT.

%============================================================
\subsubsection{Single-Precision (FP32) and Half-Precision (FP16)}

The original ERT is written in C and only supports double precision (FP64) measurements. 
While this can be easily extended to single-precision (FP32) by replacing `double' by `single' in the code, it requires work to support half-precision (FP16). 
For maintainability and future extensibility purposes, we have rewritten ERT in C++ and leverage C++ templates to support multiple data types.

\begin{table}[!htp]
\caption{FP16 Performance on CUDA Core on V100 GPUs}
\label{tab:fp16}
\centering
\begin{tabular}{|c|c|c|}
\hline
Version & Implementation & Performance (TFLOP/s)\\\hline
v1 & naive & 15.421\\\hline
v2 & replace \textit{half} with \textit{half2} & 20.142\\\hline
v3 & \textit{uint32\_t} for indexing & 28.152\\\hline
v4 & inline intermediate variables & 28.376\\\hline
v5 & \textit{uint32\_t} only & 29.182\\\hline
\end{tabular}
\end{table}

For FP32, we have easily obtained 15.2~TFLOP/s peak performance, which is within 5\% of the advertised 15.7~TFLOP/s performance~\cite{nvidia2017v100}.

For FP16 (on the CUDA Core), some performance tuning is required as detailed in Tab.~\ref{tab:fp16}. 
The naive implementation (v1) simply passes \textit{half} as the data type to the templated functions and that resulted in a similar performance to the FP32 precision's, 15.4~TFLOP/s. 
This is because V100s do not support FP16 directly on the CUDA Core~\cite{nvidia2017v100} and each FP16 operation is essentially executed as an FP32 operation (i.e. going through the same pipeline). 
To efficiently perform FP16 operations (even though utilizing the Tensor Core would be a good option), on the CUDA Core, a vector type \textit{half2} can be used to pack two FP16 values together to one FP32 register and be executed in one FP32 instruction. 
In ERT, we have implemented this using intrinsic functions and obtained an improved performance of 20.1~TFLOP/s (v2) in Tab.~\ref{tab:fp16}.
In real life, it is not feasible to implement large scale applications in intrinsics but out the implementation is an attempt to push the Roofline ceiling as high as we possibly can.

The rest three versions v3-v5 in Tab.~\ref{tab:fp16} are a series of optimizations that have proved to be beneficial to the development of ERT and are expected to be largely helpful to real-life applications and their performance tuning as well.
Out of the three, replacing \textit{uint64\_t} indexing variables with the \textit{uint32\_t} data type has proven to bring the most performance gain, from 20.1~TFLOP/s to 28.2~TFLOP/s. 
This is due to the fact that V100s only support INT32 integer operations on the hardware level and that there is constant type conversion between \textit{uint64\_t} and \textit{uint32\_t} for the second version of ERT (v2). 
With the inlining of intermediate variables in v4 and conversion of all integers to \textit{uint32\_t} in v5, the FP16 CUDA Core performance of ERT has been brought on par to the theoretical peak with 29.2~TFLOP/s in Fig.~\ref{fig:ert_roofline}.

\subsubsection{Tensor Core}
NVIDIA Tensor Cores are designed to accelerate matrix-matrix multiplication operations, which represent the mathematical nature of many deep learning workloads, for example, convolutional neural networks (CNNs). 
They operate on $4\times4$ matrices and can perform the following matrix multiplication and accumulation extremely efficiently.
\begin{equation}
\label{eq:tensor}
D = A\times B + C
\end{equation}
where A and B are matrices in FP16, and C and D are matrices in either FP16 or FP32. V100 has 80 SMs and 8 tensor cores per SM, and at 1.312 GHz clock frequency, its theoretical Tensor Core peak can be calculated as
\begin{equation}
\label{eq:tensor_peak}
80\times 8\times 1.312\times 4^{3}\times 2 = 107.479 \textrm{~TFLOP/s}
\end{equation}
To stress test the Tensor Cores on V100, we have implemented ERT based on general matrix-matrix multiplications (GEMMs), where $\alpha$ and $\beta$ are constant coefficients:
\begin{equation}
\label{eq:gemm}
D = \alpha * A\times B + \beta * C
\end{equation}

In general, there are two ways to program on Tensor Core, using the WMMA (Warp Matrix Multiply Accumulate) API in CUDA~\cite{wmma}, or libraries such as cuBLAS~\cite{cublas} and cuDNN~\cite{chetlur2014cudnn}. 
The \textit{nvcuda::wmma} namespace in CUDA provides specialized matrix load, multiply, accumulate and store operations and allows for direct programming on Tensor Cores.
cuBLAS and cuDNN libraries, on the other hand, shields users away from low-level CUDA programming and provides a very versatile, and highly-tuned, high-level user API for GEMM and other operations. 

\begin{figure}[!htp]
\centering
\includegraphics[width=0.9\columnwidth]{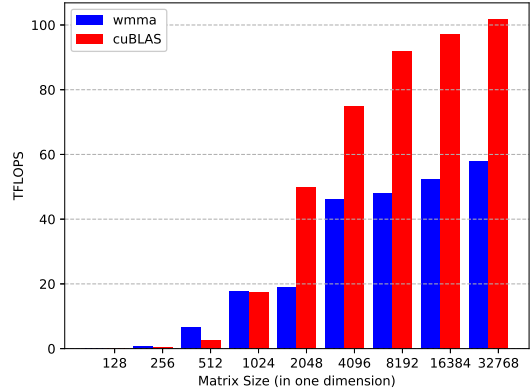}
\caption{Tensor Core Performance as a function of matrix size for cuBLAS and hand-optimized WMMA implementations of matrix multiplication.}
\label{fig:tensor_peak}
\end{figure}

For a given GEMM in Equation~\ref{eq:gemm} with matrix size $M \times N$ for $A$, $N\times K$ for $B$, and $M\times K$ for $C$ and $D$, if $M=N=K$, the total number of FLOPs performed in this kernel can be calculated as $M^{3}\times 2$. 
This is an estimation without including the constant efficiency multiplications, which usually are performed on the CUDA Core, not Tensor Core, and are negligible.
With the run time $t$, we can then estimate the FLOP/s performance of the kernel as $(M^{3}\times 2)/t$ for a given matrix size in Fig.~\ref{fig:tensor_peak}.

It is clear that as the matrix size increases, so does the performance of both \textit{wmma} and \textit{cuBLAS} approaches. 
At the largest with $M=N=K=32768$, we have obtained 103.7~TFLOP/s at 96.5\% of the theoretical peak from the \textit{cuBLAS} approach, and 58~TFLOP/s at 54\% from the \textit{wmma} approach.
This is largely due to the optimizations in \textit{cuBLAS} such as the use of shared memory, data padding (to avoid bank conflicts in shared memory), highly tuned thread block size, tile size, and other parameters. 

For the rest of this paper, we will use 103.7~TFLOP/s as the Tensor Core peak; however, the 58~TFLOP/s performance provides an empirical upper bound for users who program in \textit{wmma} on the Tensor Core.

%============================================================
\subsection{Nsight Compute Metrics for Application Characterization \label{sec:nsight}}
The application characterization methodology for Roofline analysis on NVIDIA GPUs has been evolving with the developer toolchain change. 
The first proposed methodology was based on \texttt{nvprof}~\cite{nvidia2019profiler} in~\cite{yang2019hierarchical}, and then an \texttt{Nsight Compute}~\cite{nvidia2019developer} based methodology is developed at~\cite{roofline2020} and briefly presented in~\cite{yang2020metrics}.
In this paper, we will discuss in detail how the \texttt{Nsight Compute} metrics can be used for hierarchical Roofline analysis on NVIDIA GPUs and demonstrate its effectiveness in analyzing deep learning applications.

The \texttt{Nsight} profiling toolkit is replacing \texttt{nvprof} as the new performance tool suite for NVIDIA GPU developers. 
It consists of three components, \texttt{Nsight Systems}, \texttt{Nsight Compute}, and \texttt{Nsight Graphics}, with the first two being most relevant to scientific application and machine learning application development. 
\texttt{Nsight Systems} can provide 
a system-wide visualization of application performance and help users identify issues such as insufficient parallelism on the GPU, unnecessary device-host data transfers, and inefficient kernel synchronization, while \texttt{Nsight Compute} dives a bit deeper and allows for the collection of more detailed performance metrics such as warp issues statistics, instruction pipeline utilization, and memory access pattern. 

Between the two generations of developer tools, \texttt{nvprof} and \texttt{Nsight Compute} have a few major differences. 
\begin{itemize}
    \item \texttt{nvprof} uses CUPTI~\cite{cupti} while \texttt{Nsight Compute} is based on PerfWorks~\cite{perfworks}, a new framework for performance metric collection.
    \item The metrics in \texttt{Nsight Compute} are more nuanced than in \texttt{nvprof}, with some metrics broken down into more in \texttt{Nsight Compute}.
    \item The naming and organizing convention in \texttt{Nsight Compute} is more structured as well, with components such as unit, subunit, interface, counter name, rollup metric and submetric, used to distinguish different metrics.
    \item Kernel replay when multiple metrics are being collected, and profiling overhead, are more optimized in \texttt{Nsight Compute}, to provide faster and more accurate hardware and software counter measurements.
\end{itemize}

To construct a hierarchical Roofline on NVIDIA GPUs, we need to collect the following quantities, kernel run time, the total number of FLOPs performed in each kernel, and the number of bytes being read and written at each level of the memory hierarchy.
With \texttt{Nsight Compute}, we can use this command to collect metrics listed in Tab.~\ref{tab:mapping}.

% Kernel run time, the total number of FLOPs and bytes read and written by each level of the memory hierarchy are three crucial datasets required by Roofline analysis. We use the command shown below to collect these data:
{

\centering
\textit{nv-nsight-cu-cli -{}-metrics \textbf{metric} ./application}\\\vspace{1mm}
}

\subsubsection{Kernel Run Time}
As shown in TABLE~\ref{tab:mapping}, we use the metric \texttt{sm\_\_cycles\_elapsed.avg} to obtain the total number of elapsed \texttt{cycles} and its submetric \texttt{per\_second} to get the \texttt{rate} (number of cycles per second), in order to calculate the kernel execution time:\vspace{1mm}
\begin{equation}
\label{eq:roofline_time}
time = cycles / rate
\end{equation}

%============================================================
\subsubsection{FLOPs}
To count the number of FLOPs performed in the kernel, \texttt{Nsight Compute} doesn't provide a unified metric like \texttt{flop\_count\_dp} in \texttt{nvprof}. But for each floating-point precision (FP64, FP32 and FP16), it splits the measurement into three metrics based on the instruction type, addition, multiplication, and fused multiply-add (FMA). Note that each FMA is considered two FLOPs and the total number of FLOPs can be calculated as \texttt{add + 2 x fma + mul} for each data precision.
Also, one can tell from the naming of the metrics that only non-predicated threads are counted in these FLOPs, i.e. masked operations are not included. 

For Tensor Core, we count the number of warp instructions 
% $Inst_{tc}$ 
by using the \texttt{sm\_\_inst\_executed\_pipe\_tensor.sum} metric and the total Tensor Core FLOPs is
\begin{equation}
\label{eq:roofline_tcflop}
FLOP_{tc} = Inst_{tc}\times 512
\end{equation}

%============================================================
\subsubsection{Bytes}
Metrics are listed in TABLE~\ref{tab:mapping} for measuring the data movement on each level of the memory hierarchy. 

\begin{table}[htb]
\caption{Nsight Compute metrics for hierarchical Roofline }
\label{tab:mapping}
  %\centering
 \resizebox{\columnwidth}{!}{
  \begin{tabular}{{|c|c|}}
  \hline
    & \textbf{Metrics}\\
  \hline\hline
\multirow{2}{*}{Time}
    & sm\_\_cycles\_elapsed.avg\\
    & sm\_\_cycles\_elapsed.avg.per\_second\\
\hline\hline
\multirow{3}{*}{FP64 FLOPs}
    & sm\_\_sass\_thread\_inst\_executed\_op\_hadd\_pred\_on.sum\\
    & sm\_\_sass\_thread\_inst\_executed\_op\_hmul\_pred\_on.sum\\
    & sm\_\_sass\_thread\_inst\_executed\_op\_hfma\_pred\_on.sum\\
    \hline
\multirow{3}{*}{FP32 FLOPs}
    & sm\_\_sass\_thread\_inst\_executed\_op\_fadd\_pred\_on.sum\\
    & sm\_\_sass\_thread\_inst\_executed\_op\_fmul\_pred\_on.sum\\
    & sm\_\_sass\_thread\_inst\_executed\_op\_ffma\_pred\_on.sum \\
\hline
\multirow{3}{*}{FP16 FLOPs}
    & sm\_\_sass\_thread\_inst\_executed\_op\_hadd\_pred\_on.sum\\
    & sm\_\_sass\_thread\_inst\_executed\_op\_hmul\_pred\_on.sum\\
    & sm\_\_sass\_thread\_inst\_executed\_op\_hfma\_pred\_on.sum\\
\hline
    Tensor Core FLOPs & sm\_\_inst\_executed\_pipe\_tensor.sum\\
\hline\hline
\multirow{1}{*}{L1 Cache}
         & l1tex\_\_t\_bytes.sum \\
\hline
\multirow{1}{*}{L2 Cache}
    & lts\_\_t\_bytes.sum \\
\hline
\multirow{1}{*}{HBM}
    & dram\_\_bytes.sum\\
\hline
  \end{tabular}
  }
\end{table}

For device memory (or HBM), L2 cache, and L1 cache, the latest \texttt{Nsight Compute} provides a unified byte metric for each of them to facilitate measurement. Note that shared memory transactions are not included in the current L1 metric.

Due to profiling overhead, it is recommended to restrict the number of kernels to run \texttt{Nsight Compute} with at a time, and these metrics can be collected on separate runs as well, as long as the execution of the application is deterministic. 
Also, note that as of 2020.1.0, \texttt{Nsight Compute} serializes multi-stream execution so certain performance gain due to kernel overlapping may be overlooked; however, the performance analysis in this paper is still insightful in understanding application performance on a kernel level.
% It should be noted that the current methodology can be only applied to a single-GPU test case since \texttt{Nsight Compute} will cause deadlock in NVIDIA Collective Communications Library (NCCL) kernels. Besides, \texttt{Nsight Compute} will serialize multi-stream execution on NVIDIA GPUs so one cannot get the accurate overall run time by simply summing up all kernel run time together.
% Note that all metrics can be collected in one run or separately, one at a time. It is recommended to restrict the number of kernels to profile to reduce overall overhead.

%============================================================
%\subsection{Time-Based Roofline}

% %============================================================
\section{Experimental Setup}
\subsection{Hardware and Software Configuration}
Results presented in this paper are obtained from the Cori supercomputer, and in particular its GPU partition, at the National Energy Research Scientific Computing Center (NERSC), Lawrence Berkeley National Laboratory (LBNL). 
The GPU partition is primarily deployed for GPU porting, benchmarking, and testing efforts in the NERSC Exascale Science Application Program (NESAP). 
Each node contains two Intel Xeon Gold 6148 Skylake CPUs, 384GiB DDR4 memory, and 8 NVIDIA V100 GPUs.
Each GPU has 16GiB of HBM2 memory and 80 SMs, and GPUs on a node are connected to each other in a `hybrid cube-mesh' topology.

On the software side, we have used the TensorFlow 1 and PyTorch implementation of the climate image segmentation code in~\cite{deepcam}, and CUDA 10.2.89, cuDNN 7.6.5, Nsight Compute 2020.1.0, Python 3.7, PyTorch 1.5.0, and TensorFlow 1.15.0 for this study.

\subsection{DeepCAM Benchmark}
% \fix{maybe ask Steve/Thorsten to write a short description of the app, and whether/how the TF and PT versions differ in implementation}
DeepCAM\cite{deepcam} is a deep learning benchmark extracted from the 2018 Gordon Bell winning project~\cite{kurth2018exascale}, used for detection, classification, and localization of extreme weather patterns in climate images.
It has two different implementations, in TensorFlow and PyTorch respectively, with the PyTorch version being selected for MLPerf~\cite{mlperf} HPC benchmark suite. 
In this paper, we will compare the performance of these two implementations using the methodology presented in Sec.~\ref{sec:nsight}.
To ensure a fair comparison, we have tuned the parameters to be as close as possible, for example, the number of layers in the encoder-decoder architecture, layer parameters, optimization algorithms, step rates, batch size, usage of batch norm, and Automatic Mixed Precision (AMP) settings.

% real-life application dealing with detection, classification, and localization of extreme weather patterns. It's a reference implementation for the climate segmentation benchmark, based on the exascale deep learning for climate analytics~\cite{kurth2018exascale} work. By applying a number of optimizations from both the algorithm side and the system-level side, it is the first exascale deep learning application. This benchmark has also been selected as one of the two benchmarks in the MLPerf HPC test suite. 

% In this work, we will evaluate the performance of this benchmark by comparing its two alternative implementations: one~\cite{climseg2020} is implemented in TensorFlow and another~\cite{deepcam} is implemented in PyTorch. Both frameworks take single-precision data as input and use NVIDIA Automatic Mixed Precision (AMP) to accelerate the training process. Besides, the TensorFlow version also incorporates a half-precision mode which is a manual effort of AMP.

% \fix{more details about TF and PT implementations...}
% % - introduce the encoder-decoder architecture in TF, and how it's written in PT

The DeepCAM model is a deep neural network for semantic segmentation with an encoder-decoder architecture based on DeepLabv3+~\cite{Chen_2018_ECCV}. The encoder is a ResNet-50 network with atrous spatial pyramid pooling. The decoder is a nine-layer network with convolutional and de-convolutional layers and two skip connections from the input and middle of the encoder.

To profile the code, the \texttt{profile-from-start} option is disabled in \texttt{Nsight Compute} and we use CuPy~\cite{okuta2017cupy} to explicitly restrict the profiling region to include the iteration loop only. 
To have relatively stable run time behavior during profiling, we also set up a warm-up loop with 5 iterations before the target profiling loop. We collect only one metric during each execution to minimize the profiling overhead which will result in random algorithmic choices due to the TensorFlow runtime auto-tuning. To solve this issue, NVIDIA TensorFlow Determinism~\cite{determinism} is employed to get rid of this uncertainty.

If not otherwise stated, the default setting for the TensorFlow DeepCAM implementation is with AMP-enabled, and for PyTorch DeepCAM with AMP optimization level \texttt{O1}.
The source code and full raw results are available at~\cite{deepcam}.

% Similar settings are used to make a fair comparison between these two frameworks. Unless otherwise stated, batch normalization is enabled by default and the batch size is 2. The PyTorch AMP optimization level is set as \texttt{O1}. As for profiling, the \texttt{profile-from-start} option is disabled in \texttt{Nsight Compute} and we use CuPy~\cite{okuta2017cupy} to explicitly restrict the profiling region to include the iteration loop only. To have relatively stable run time behavior during profiling, we also set up a warm-up loop with 5 iterations before the target profiling loop. We collect only one metric during each execution to minimize the profiling overhead which will result in random algorithmic choices due to the TensorFlow runtime auto-tuning. To solve this issue, NVIDIA TensorFlow Determinism~\cite{determinism} is employed to get rid of this uncertainty.
%- we tried to do apples to apples comparison:
%same/similar number of layers, layer parameters, opt algorithm (Larc Adam), training/test dataset, batch size, batchnorm, AMP, etc
%- used the determinism package  to pick a fixed algorithm for autotuning, avoid underlying algorithm change when profiling/collecting different metrics; direct readers to see details in that repo if they're interested
%source code and raw results are available at~\cite{resdeepcam}

%============================================================
\section{Results}
In this section, we will first apply the \texttt{Nsight Compute} methodology in Sec.~\ref{sec:nsight} on the DeepCAM benchmark and discuss its performance implications. On the following Roofline charts, each kernel is represented by a triplet of open circles (blue for L1, red for L2 and green for HBM), and the circle size is proportional to the kernel's run time. Note that we preset a minimum circle size to make all kernels visible on the plot, and that the real run time difference between large and small kernels can be more significant.
Besides, there could be many invocations of the same kernel and the data presented on these Roofline charts is the aggregation of all these invocations of the same kernel.
One should expect blue, red, and green circles near the L1, L2, and HBM ceilings respectively to show high memory utilization.
Triplets of circles close to each other present a ``streaming'' data access pattern and indicate poor cache locality. Circles to the top right corner show superior performance over the others.
%\fix{I think there needs to be some high-level guidance like look for red dots between L1 and L2 bandwidth, blue dots between HBM and L2, or green dots near the HBM line... triplets of circles near each other are streaming... want circles to be to the right and near the top(TC)}

In the following subsections, we will discuss how performance is different in the forward and backward pass in both TensorFlow and PyTorch implementations, and the performance impact of the NVIDIA Automatic Mixed Precision package and the zero-AI kernels.
Note that the backward pass for TensorFlow DeepCAM includes both gradient calculation and gradient update, whereas the PyTorch DeepCAM backward pass only includes gradient calculation (with its `optimizer' being the gradient update step).

% In this section, we discuss our observations and performance insights by applying \texttt{Nsight Compute} methodology on the DeepCAM benchmark. Computation kernels are presented by open circles in the hierarchical Roofline and circle dimensions denote their run time. The larger the circle is, the longer run time it requires. For each framework, we focus on the performance comparison between the forward and backward pass and the performance enhancement brought by NVIDIA Automatic Mixed Precision.

\subsection{The TensorFlow version of DeepCAM}
\begin{figure}[!htp]
\centering
\includegraphics[width=0.9\linewidth]{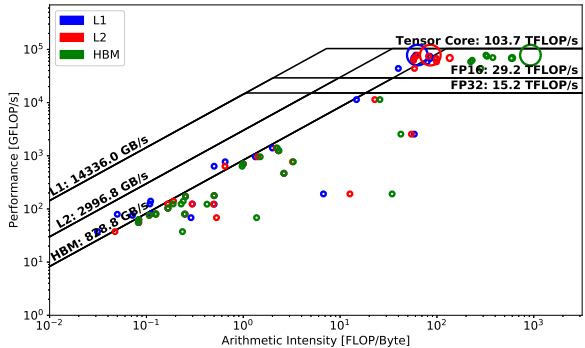}
\caption{Hierarchical Roofline of the TensorFlow DeepCAM in the forward pass with default configurations. The dominant kernel (with three largest circles) has very high Tensor Core utilization and consume 33\% of the overall run time.}
\label{fig:tf_fw}
\end{figure}
Fig.~\ref{fig:tf_fw} shows the hierarchical Roofline of the TensorFlow version of DeepCAM in its forward pass. The main computational kernel represented by the three large circles under the Tensor Core ceiling, indicates that it has very high Tensor Core utilization, whereas many of the other circles either do not use Tensor Core or are bandwidth bound.
This major kernel's L1 circle (in blue) slightly overlaps with its L2 circle (in red) indicating a relatively low L1 cache locality; however, the large gap between its L2 and HBM circles demonstrates that L2 cache misses rarely happened and that the kernel benefits from high L2 data locality. As for the rest of the kernels, their L1, L2, and HBM kernels are generally close to each other, implying a poor data locality across all levels of memory hierarchies (``streaming'' operations).
% Fig.~\ref{fig:tf_fw} shows the hierarchical Roofline of the TensorFlow version of DeepCAM in its forward pass with default configurations. The major computation kernel represented by the large circles right under the Tensor Core ceiling indicates a very high Tensor Core utilization. Its L1 circle (in blue) slightly overlaps with its L2 circle (in red) indicates a relatively low L1 cache locality. The huge gap between L2 and HBM circles, on the other hand, demonstrates that L2 cache misses rarely happened and the kernel benefits from high L2 data locality. As for the rest trivial kernels, their L1, L2, and HBM kernels are generally close to each other implying a poor data locality across all levels of memory hierarchies.
\begin{figure}[!htp]
\centering
\includegraphics[width=0.9\linewidth]{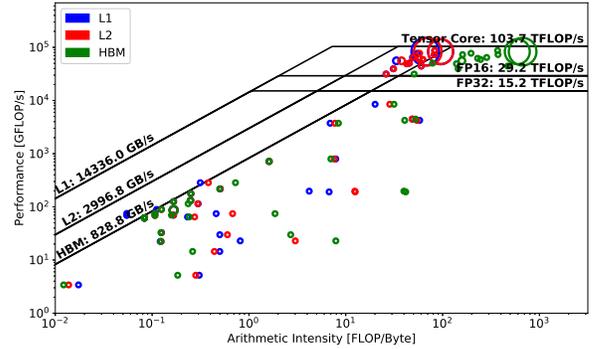}
\caption{Hierarchical Roofline of the TensorFlow DeepCAM in the backward pass with default configurations. There are more compute-intensive kernels than in the forward pass. Collectively they constitute 41.9\% of the run time and attain near peak Tensor Core performance.}
\label{fig:tf_bw}
\end{figure}

Fig.~\ref{fig:tf_bw} shows the corresponding backward pass of the TensorFlow DeepCAM. Instead of one single major kernel appearing in the forward pass, two very time-consuming kernels are found in the backward pass calculation. It is obvious that these two kernels both require longer run time than the major kernel in the forward pass (notice the size), which implies that the backward pass has more compute-intensive kernels than the forward pass and is generally more time-consuming. Compared to a few kernels using Tensor Core in the forward pass, we can find that more kernels benefit from the Tensor Core pipeline in the backward pass since they are sitting above the half-precision peak. Another observation is that more kernel invocations are involved in the backward pass than in the forward. Overall, we can conclude that in either forward or backward pass, the main computational kernels are compute-bound and are highly optimized for the underlying architecture.

\subsection{The PyTorch version of DeepCAM}
\begin{figure}[!htp]
\centering
\includegraphics[width=0.9\linewidth]{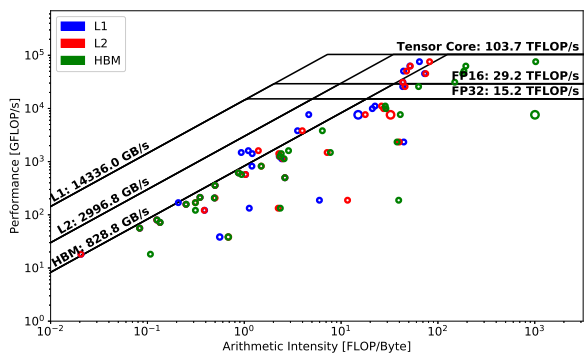}
\caption{Hierarchical Roofline of the PyTorch DeepCAM in the forward pass with default configurations. No single kernel requires significantly longer run time than the others (no extremely large circles).}
\label{fig:pt_fw}
\end{figure}
Compared to the TensorFlow result (Fig.~\ref{fig:tf_fw}), no dominant kernels (kernel run time significantly larger than the others) can be found in the PyTorch forward pass (Fig.~\ref{fig:pt_fw}). The number one kernel is located slightly below the single-precision performance peak, and based on the symbol distance between different memory hierarchies, it has a better cache utilization than the dominant kernel in TensorFlow (even though it runs on the CUDA Core). Besides, similar to TensorFlow, a large number of trivial kernels are HBM-bound in the PyTorch implementation of DeepCAM. 
\begin{figure}[!htp]
\centering
\includegraphics[width=0.9\linewidth]{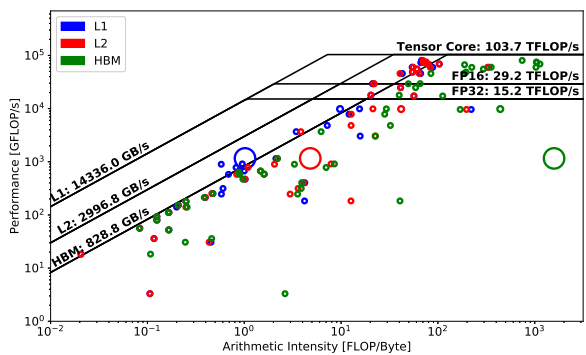}
\caption{Hierarchical Roofline of the PyTorch DeepCAM in its backward pass with default configurations. One can observe the highly compute intensive, but low performing kernel.}
\label{fig:pt_bw}
\end{figure}

Fig.~\ref{fig:pt_bw} shows the PyTorch DeepCAM performance in the backward pass, with default configurations. Surprisingly, the number one time-consuming kernel does not utilize Tensor Core and delivers only about 1~TFLOP/s performance. However, this implementation's overall run time is still lower than that of the TensorFlow case, seen by the size of the circles, thanks to optimizations in other kernels or the overall execution of kernels.

\begin{figure}[!htp]
\centering
\includegraphics[width=0.9\linewidth]{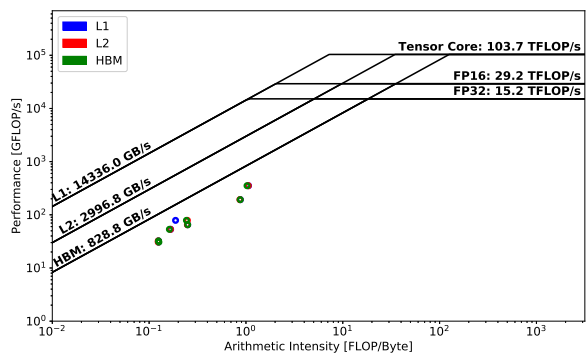}
\caption{Hierarchical Roofline of the PyTorch DeepCAM in its `optimizer' step. The gradient update step consists of numerous streaming operations and has poor arithmetic intensity and FLOP/s performance.}
\label{fig:pt_opt}
\end{figure}

Compared to TensorFlow, PyTorch has more flexibility when profiling the model, and the `optimizer' step can be easily separated from the gradient calculation step in the back-propagation. The optimization step is mainly to update model parameters with newly calculated gradients and is usually low on arithmetic intensity.
Fig.~\ref{fig:pt_opt} confirms this, where all the `optimizer' kernels are memory-bound and have a much lower FLOP/s performance than some of the kernels in Fig.~\ref{fig:pt_fw} or Fig.~\ref{fig:pt_bw}. It should be noted that there are 2709 kernel invocations involved in this process, even though there are only a few circles visible. These kernel invocations have very similar arithmetic intensity and performance, and are thus overlapping.

\subsection{Automatic Mixed Precision}
\begin{figure}[!htp]
\centering
\includegraphics[width=0.9\linewidth]{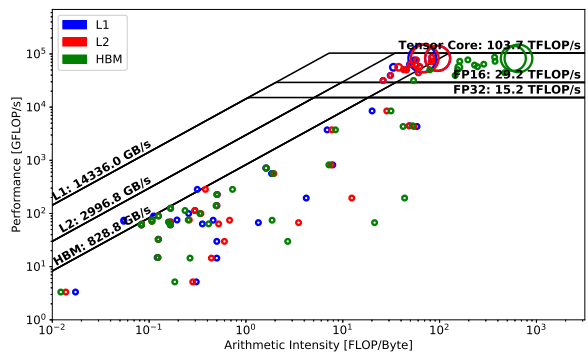}
\caption{Hierarchical Roofline of our FP16 implementation of DeepCAM in TensorFlow (backward pass). AMP (shown in Fig.~\ref{fig:tf_bw}) can deliver the same performance without manual type conversion.}
\label{fig:tf_bw_fp16}
\end{figure}
The Automatic Mixed Precision (AMP) package developed at NVIDIA is dedicated to accelerating deep learning processes by partially converting single-precision data to half-precision to reduce data movement and improve computational throughput. 
It allows for automatic type conversion of certain model parameters and also implements schemes such as loss scaling to ensure numerical correctness and accuracy.
% It is different from the naive type casting since certain sensitive operations like \texttt{softmax} are performed in single-precision only. 
% Moreover, the AMP conversion is aimed to minimize numerical errors like round-offs so the numerical accuracy can be guaranteed. 
We have implemented an FP16 version of DeepCAM in TensorFlow manually, by picking out the appropriate variables by hand and typecasting them explicitly.
Fig.~\ref{fig:tf_bw_fp16} shows that the backward pass performance of this implementation is very close to that of the FP32 DeepCAM with AMP-enabled (shown in Fig.~\ref{fig:tf_bw}), demonstrating that even without the knowledge of the implementation details of the network, the AMP package can effectively apply type conversion and leverage lower-precision operations for performance.
%\fix{I don't follow what was done here... sounded like TF DeepCAM vs. TF+AMP DeepCAM delivered about the same performance.  Why would it be different?  Why not start with 32b TF DeepCAM and show AMP can selectively find the best kernels for FP16?}

% of the FP16 TensorFlow implementation which is a manual effort of the default FP32 AMP code (performance shown in Fig.~\ref{fig:tf_bw}). Results show that few differences can be identified from these two implementations which implies any manual efforts to explore mixed-precision acceleration might not be worthwhile since AMP is already performing well.
\begin{figure}[!htp]
\centering
\includegraphics[width=0.9\linewidth]{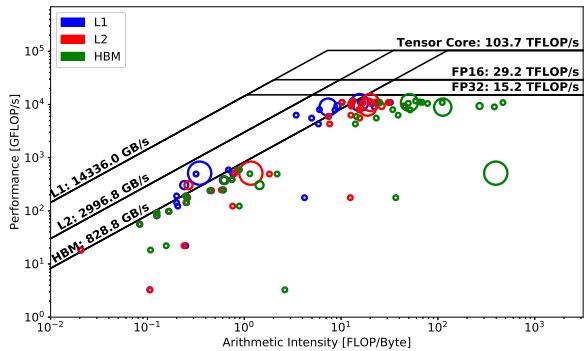}
\caption{Hierarchical Roofline of the PyTorch DeepCAM in its backward pass with AMP O0.}
\label{fig:pt_bw_o0}
\end{figure}

AMP provides implementation for both TensorFlow and PyTorch, and for PyTorch, there are more detailed optimization levels, rather than just on or off.
% Unlike with TensorFlow, AMP provides more detailed optimization levels for PyTorch.
% where one can only choose between with or without AMP, users can further set different AMP optimization levels in PyTorch. 
According to the AMP documentation~\cite{amp2020}, \texttt{O0} level for PyTorch is used to establish a stable baseline for the auto mixed-precision acceleration; \texttt{O1} follows a conservative type conversion and numerical properties are highly preserved; \texttt{O2} however, implements a more aggressive FP32 to FP16 conversion and extra care needs to be taken for model convergence concerns.
% thus it's supposed to deliver higher performance than \texttt{O1}. 
% We use the default \texttt{O1} level throughout this work and the performance difference between \texttt{O1} and \texttt{O2} is invisible in our tests. 

Our default setting is \texttt{O1} and the backward pass performance of the PyTorch DeepCAM with this setting is shown in Fig.~\ref{fig:pt_bw}.
From the \texttt{O0} optimization level in Fig.~\ref{fig:pt_bw_o0}, to the \texttt{O1} in Fig.~\ref{fig:pt_bw}, kernel run time has been largely reduced and many kernels have been moved to execute on the Tensor Core, providing a much higher computational throughput and demonstrating the effectiveness of the \texttt{O1} optimization level.

% Fig.~\ref{fig:pt_bw_o0} shows the PyTorch backward pass with \texttt{O0} optimization level. Compared to 
% Fig.~\ref{fig:pt_bw}, we can find that the largest performance bottleneck is slightly shifted to the right (arithmetic intensity is improved) with \texttt{O1} optimization. Also, with \texttt{O1} optimization, a number of kernels which was limited by the single-precision ceiling (kernels closed to the FP32 peak in Fig.~\ref{fig:pt_bw_o0}) can benefit from the Tensor Core pipeline thus their run time becomes trivial for the whole computation.

\subsection{Zero-AI kernels}
% \begin{figure}[!htp]
% \centering
% \includegraphics[width=0.7\linewidth]{figures/tf-invoc.pdf}
% \includegraphics[width=0.7\linewidth]{figures/pt-invoc.pdf}
% \caption{Ratio of zero-AI kernel invocations to the total number of kernel invocations in two implementations, TensorFlow DeepCAM and PyTorch DeepCAM. Ratios are similar, but counts are very different.}
% %what about aggregate time in zero- and non-zero AI kernels?}}
% \label{fig:tf_invoc}
% \end{figure}
% \begin{figure}[!htp]
% \centering
% \includegraphics[width=0.7\linewidth]{figures/pt-invoc.pdf}
% \caption{Ratio of zero-AI kernel invocations to the total number of kernel invocations in PyTorch DeepCAM.}
% \label{fig:pt_invoc}
% \end{figure}

Compared to traditional HPC applications where users usually have full control of kernel invocations, high-level Python-based deep learning frameworks tend to implicitly invoke many subsidiary kernels, either for data conversion or device-host transfer purposes. 
TABLE~\ref{tab:0-ai} shows the ratio of these kernel invocations to the total number of invocations. 
Around 40-50\% of the invocations are for such zero-AI kernels, where no floating-point operation is performed.
This may not inadvertently affect the overall performance much if these kernels are perfectly overlapped with other kernel executions, but it is very hard to achieve that in reality. %\fix{but they could be giant/slow mem copies.}
As hardware constantly evolves, new computer architectures tend to provide higher and higher FLOP/s performance and bandwidth, but with less progressive improvement on kernel launch overhead. 
To avoid becoming overhead-bound, it is recommended that these deep learning applications avoid such ``implicit" zero-AI kernels as much as possible by fusing them or overlapping with the non-zero-AI kernels.

\begin{table}[!htp]
\caption{Zero-AI kernel invocations in TensorFlow DeepCAM and PyTorch DeepCAM}
\label{tab:0-ai}
\resizebox{\columnwidth}{!}{
  \begin{tabular}{{|c|c|c|c|c|}}
  \hline
  TensorFlow DeepCAM & Forward & \multicolumn{2}{c|}{Backward$^{\mathrm{a}}$} & Total \\
  \hline
  zero-AI & 304 (54.7\%) & \multicolumn{2}{c|}{1833 (40.1\%)} & 2137 \\
  \hline
  non zero-AI & 252 (45.3\%) & \multicolumn{2}{c|}{2740 (59.9\%)} & 2992\\
  \hline
  Total & 556 (100\%) & \multicolumn{2}{c|}{4573 (100\%)} & 5129\\
  \hline\hline
  PyTorch DeepCAM & Forward & Backward & Optimizer & Total \\
    \hline
  zero-AI & 437 (54.8\%) & 609 (38.7\%) & 0 (0\%) & 1046 \\
  \hline
  non zero-AI & 360 (45.2\%) & 966 (61.3\%) & 2709 (100\%) & 4035 \\
  \hline
  Total & 797 (100\%) & 1557 (100\%) & 2709 (100\%) & 5081\\
  \hline
  \multicolumn{5}{l}{$^{\mathrm{a}}$This includes both gradient calculation and update, i.e. the backward pass and}\\
  \multicolumn{5}{l}{optimizer in the PyTorch case.}
  \end{tabular}
 }
\end{table}

\subsection{Overall Performance}

Despite minor differences in implementation (even though we have tried to make an apples-to-apples comparison), the two codes, TensorFlow DeepCAM and PyTorch DeepCAM, have achived similar runtime and convergence performance. 
The previous subsections presented a deep analysis of these two implementations on hierarchical Roofline, and it is discovered that TensorFlow tends to utilize Tensor Core more, compared to PyTorch, as seen by the locations of the most time-consuming kernels in Fig.~\ref{fig:tf_fw}-\ref{fig:pt_bw}. 
These two frameworks have similar cache utilization pattern on L1, L2 and HBM levels, with PyTorch having slightly more high-AI kernels scattered in the range of 100 FLOPs/Byte and 1000 FLOPs/Byte on Fig.~\ref{fig:pt_fw} and Fig.~\ref{fig:pt_bw}.

Overall, similar numbers of kernels are launched in TensorFlow DeepCAM and PyTorch DeepCAM, with TensorFlow using over double the amount of zero-AI kernels than in PyTorch, 2137 versus 1046 in Tab.~\ref{tab:0-ai}. 
These zero-AI kernels may have been launched over multiple streams and overlapped with computational kernels, however, reducing them could further improve the launch overhead and overall run time. 
These kernels are mostly used for converting data from one precision to another, or for rearranging data layout. 
They may be fused or done on the host (asynchronous to the GPU computation) in order to save run time.

Another note is that the NVIDIA AMP package has been proven to be very effective, through the comparison of Fig.~\ref{fig:tf_bw} and Fig.~\ref{fig:tf_bw_fp16} for TensorFlow, and Fig.~\ref{fig:pt_bw} and Fig.~\ref{fig:pt_bw_o0} for PyTorch.

\section{Conclusions}
In this paper, we first revisited the need for mixed-precision performance analysis and extended ERT to incorporate single-precision, half-precision, and Tensor Core performance measurements. Then, based on the previous \texttt{nvprof} hierarchical Roofline methodology, we established a new \texttt{Nsight Compute} methodology to collect Roofline data on NVIDIA GPUs. In the third part of this paper, we applied this new methodology to a representative real-life deep learning benchmark, DeepCAM, with its two implementations in TensorFlow and PyTorch. Results show that this new methodology is very effective in analyzing and better understanding the performance of deep learning applications. Useful performance insights are discussed, for example, computational characteristics of different stages of the training process, the performance impact of the automatic mixed precision (AMP) package and zero-AI kernels.
This should be largely helpful to deep learning programmers and framework developers, as it captures data localities within each level of the cache hierarchy, demonstrates overall hardware utilization and indicates potential optimization efforts (get rid of zero-AI kernels to minimize kernel launch latency and improve overall FLOP rate).
%\fix{articulate how it would be helpful.  How would a DL prog, developer, or architect use this information?}

% largely facilitates the profiling process compared to previous methods. Data collection is quick and accurate due to new performance metrics in \texttt{Nsight Compute}. We also identified that zero-AI kernels may make common deep learning applications overhead-bound and this could become the major performance bottleneck for future deep learning applications.

In the future, we would like to extend the current \texttt{Nsight Compute} methodology to incorporate cross-node performance analysis. New methodologies for alternate architectures and mixed-precision performance ceilings in Roofline will be investigated as well.
%\fix{alternate architectures?  tf16/tf32/bfloat16?}

%============================================================
\section*{Acknowledgements}
This material is based upon work supported by the Advanced Scientific Computing Research Program in the U.S. Department of Energy, Office of Science, under Award Number DE-AC02-05CH11231.
This research used resources of the National Energy Research Scientific Computing Center (NERSC) which is supported by the Office of Science of the U.S. Department of Energy under Contract No. DE-AC02-05CH11231.
We thank NVIDIA Corporation for their willingness to answer our myriad of questions on Nsight metrics.

%============================================================
\bibliographystyle{IEEEtran}
\bibliography{IEEEabrv,bibfile}

\end{document}